\documentclass[runningheads,a4paper]{llncs}
\usepackage{color}
\usepackage{amssymb}
\usepackage{cite}
\usepackage{graphicx}
\usepackage{paralist}
\usepackage{url}
\usepackage{subfig}
\usepackage{array}
\newcolumntype{L}[1]{>{\raggedright\let\newline\\\arraybackslash\hspace{0pt}}m{#1}}
\newcolumntype{C}[1]{>{\centering\let\newline\\\arraybackslash\hspace{0pt}}m{#1}}
\newcolumntype{R}[1]{>{\raggedleft\let\newline\\\arraybackslash\hspace{0pt}}m{#1}}
\usepackage{tabularx}
\usepackage{booktabs}

\usepackage{url}
\urldef{\mailsa}\path|alireza@eecs.ucf.edu, gitars@eecs.ucf.edu, klakkara@sandia.gov|
\newcommand{\keywords}[1]{\par\addvspace\baselineskip
\noindent\keywordname\enspace\ignorespaces#1}
\newcommand{\name}{}
\def\name/{RPM}

\begin{document}

\mainmatter  

\title{Leveraging Network Dynamics\\ for Improved Link Prediction}
\author{Alireza Hajibagheri\inst{1} \and Gita Sukthankar\inst{1} \and Kiran Lakkaraju\inst{2}}

\authorrunning{Hajibagheri et al.}

%
%
%

\institute{
	\mbox{}\inst{1}University of Central Florida, Orlando, Florida\\
	\inst{2}Sandia National Labs, Albuquerque, New Mexico\\
\mailsa\\
}

%
%

\maketitle

\begin{abstract}
	The aim of link prediction is to forecast connections that are most likely to occur in the future, based on examples of previously observed links.  A key insight is that it is useful to explicitly model \textit{network dynamics}, how frequently links are created or destroyed when doing link prediction.  In this paper, we introduce a new supervised link prediction framework, \name/ (Rate Prediction Model).  In addition to network similarity measures,  \name/ uses the predicted rate of link modifications, modeled using time series data; it is implemented in Spark-ML and trained with the original link distribution, rather than a small balanced subset.   We compare the use of this network dynamics model to directly creating time series of network similarity measures.  Our experiments show that \name/, which leverages predicted rates, outperforms the use of network similarity measures, either individually or within a time series.
\end{abstract}

\section{Introduction}

Many social networks are constantly in flux, with new edges and vertices being added or deleted daily.  Fully modeling the dynamics that drive the evolution of a social network is a complex problem, due to the large number of individual and dyadic factors associated with link formation.  Here we focus on predicting one crucial variable--the \textit{rate} of network change.   Not only do different networks change at different rates, but individuals within a network can have disparate tempos of social interaction.  This paper describes how modeling this aspect of network dynamics can ameliorate performance on link prediction tasks.

Link prediction approaches commonly rely on measuring topological similarity between unconnected nodes~\cite{al2011survey,getoor2005link,wang2007local}.  It is a task well suited for supervised binary classification since it is easy to create a labeled dataset of node pairs; however, the datasets tend to be extremely unbalanced with a preponderance of negative examples where links were not formed.  Topological metrics are used to score node pairs at time $t$ in order to predict whether a link will occur at a later time $t' (t' > t)$.  However, even though these metrics are good indicators of future network connections, they are less accurate at predicting \textit{when} the changes will occur (the exact value of $t'$).  To overcome this limitation, we explicitly learn link formation rates
for all nodes in the network; first, a time series is constructed for each node pair from historic data and then a forecasting model is applied to predict future values.  The output of the forecasting model is used to augment topological similarity metrics within a supervised link prediction framework.  Prior work has demonstrated the general utility of modeling time for link prediction (e.g., \cite{huang2009time,berlingerio2009,potgieter2009}); our results show that our specific method of rate modeling outperforms the use of other types of time series.

\name/ is implemented using Spark MLlib machine learning library. Using Spark, a general purpose cluster computing system, enables us to train our supervised classifiers with the full data distribution, rather than utilizing a small balanced subset, while still scaling to larger datasets.  Moreover, we evaluate the classifiers with a full test dataset, so the results are representative of the performance of the method "in the wild".  Our experiments were conducted with a variety of datasets, in contrast to prior work on link prediction that has focused on citation or collaboration networks~\cite{Liben-Nowell}. In addition to a standard co-authorship network (hep-th arXiv~\cite{soares2012time}), we analyze the dynamics of an email network (Enron~\cite{cohen2009enron}) and two player networks from a massively multiplayer online game (Travian~\cite{hajibagheri2015conflict}).  Networks formed from different types of social processes may vary in their dynamics, but our experiments show that \name/ outperforms other standard approaches on all types of datasets.

\section{Background}

Approaches to the link prediction problem are commonly categorized as being \textit{unsupervised}~\cite{Liben-Nowell,huang2009time,LNCS2014-Xi,beigi2016signed,davoudi2016modeling} or \textit{supervised} ~\cite{hasan2006,soares2012time,Sukthankar-Xi-ASONAM2013,Lichtenwalter2010}.  In unsupervised approaches, pairs of non connected nodes are initially ranked according to a chosen similarity metric (for instance, the number of common neighbors)~\cite{lu2009role,murata2008link}. The top $k$ ranked pairs are then assigned as the predicted links.  The strength of this paradigm is that it is simple and generalizes easily to many types of data, but there are some limitations: for instance, how to \textit{a priori} select the cutoff threshold for link assignment?  Implicitly, these approaches assume that the links with the highest scores are most likely to occur and form the earliest; however this is often not the case in many dynamic networks~\cite{murata2008link}.  If the rank correlation between the selected metric and the sequence of formed links is poor, the accuracy of this approach suffers.

Supervised approaches have the advantage of being able to 1) simultaneously leverage multiple structural patterns and 2) accurately fit model parameters using training data.  In this case, the link prediction task is treated as a classification problem, in which pairs of nodes that are actually linked are assigned to class 1 (positive class), whereas the non-connected ones are assigned to class 0 (negative class).   The standard model assumes that feature vectors encapsulating the current network structure at time $t$ are used to predict links formed at $t+1$; in some sense, this model is "amnesiac", ignoring the past connection history of individual nodes.  To address this issue, our proposed method, \name/ represents the network with \textit{time series}.  A forecasting model is then used to predict the next value of the series; this value is in turn used to augment the input to the supervised learning process.

\begin{figure}
  \centering 
  \includegraphics[width=0.8\textwidth]{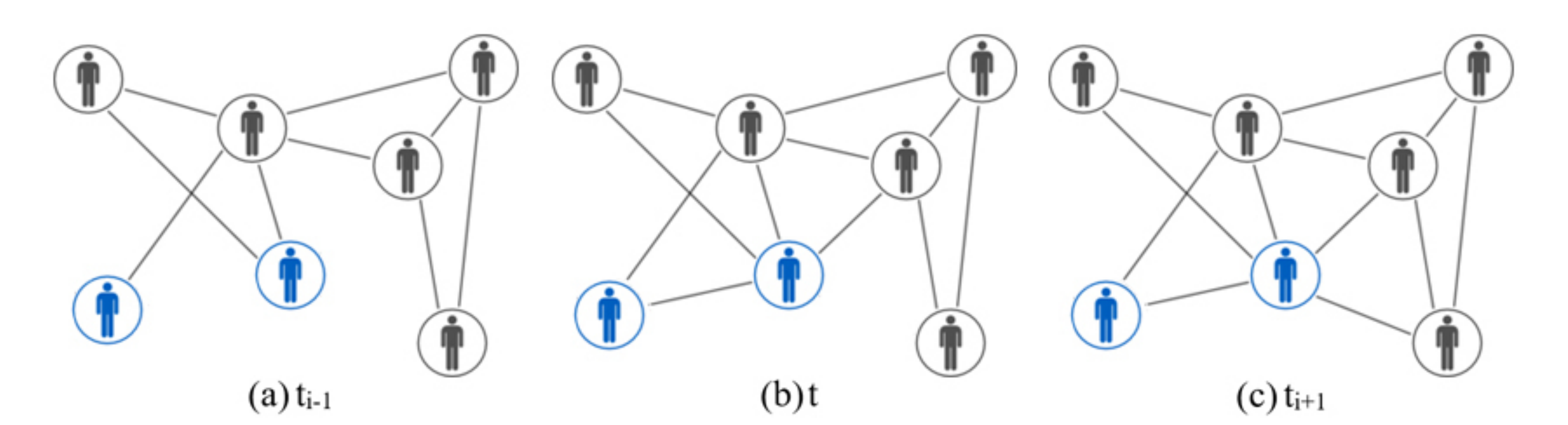}
  \caption{Evolution of a network over time.  Blue nodes have higher \textit{rates} of link formation.  This behavior can only be captured by taking temporal information into account; \name/ identifies these nodes through the use of time series.}
    \label{fig:example}
\end{figure}

\subsection{Time Series}
To construct the time series, the network $G$ observed at time $t$ must be split into several time-sliced snapshots, that is, states of the network at different times in the past. Afterwards, a window of prediction is defined, representing how further in the future we want to make the prediction. Then, consecutive snapshots are grouped in small sets called frames. Frames contain as many snapshots as the length of the window of prediction. These frames compose what is called Framed Time-Sliced Network Structure ($S$)~\cite{soares2012time}. Let $G_t$ be the graph representation of a network at time $t$. Let $[G_1,G_2,...,G_T ]$ be the frame formed by the union of the graphs from time 1 to $T$. Let $n$ be the number of periods (frames) in the series. And let $w$ be the window of prediction. Formally, $S$ can be defined as:
\[
S = \{[G_1,...,G_w],[G_{w+1},...,G_{2w}],...[G_{(n-1)w+1},...,G_{nw}]\}
\]

For instance, suppose that we observed a network from day 1 to day 9, and our aim is to predict links that will appear at day 10. In this example, the forecast horizon (window of prediction) is one day.  Our aim here is to model how the networks evolve every day in order to predict what will happen in the forecast horizon.  Figure~\ref{fig:example} shows an example of the evolution of network over time.

\subsection{Network Similarity Metrics}

In this paper, we use a standard set of topological metrics to assign scores to potential links:

\begin{compactenum}
\item Common Neighbors (CN)~\cite{Newman01clusteringand} is defined as the number of nodes with direct relationships with both members of the node pair: $CN(x,y)=|\Gamma(x)\cap\Gamma(y)|$ where $\Gamma(x)$ is the set of neighbors of node $x$. 
\item Preferential Attachment (PA)~\cite{barabasi2009scale,Liben-Nowell} assumes that the probability that a new link is created is proportional to the node degree $|\Gamma(y)|$.  Hence, nodes that currently have a high number of relationships tend to create more links in the future: $PA(x,y)=|\Gamma(x)| \times |\Gamma(y)|$. 
\item Jaccard's Coefficient (JC)~\cite{Tan-2005} assumes higher values for pairs of nodes that share a higher proportion of common neighbors relative to total number of neighbors they have: $JC(x,y)=\frac{|\Gamma(x)\cap\Gamma(y)|}{|\Gamma(x)\cup\Gamma(y)|}$. 
\item Adamic-Adar (AA)~\cite{adamic2003friends}, similar to JC, assigns a higher importance to the common neighbors that have fewer total neighbors. Hence, it measures exclusivity between a common neighbor and the evaluated pair of nodes:
	\[
	AA(x,y)=\sum_{z\in|\Gamma(x)\cap\Gamma(y)|}\frac{1}{log(|\Gamma(z)|)}.
\]
\end{compactenum}
These metrics serve as 1) unsupervised baseline methods for evaluating the performance of \name/ and 2) are also included as features used by the supervised classifiers.

\subsection{Datasets}

\begin{figure}[h!]
 \centering
\includegraphics[width=0.35\textwidth]{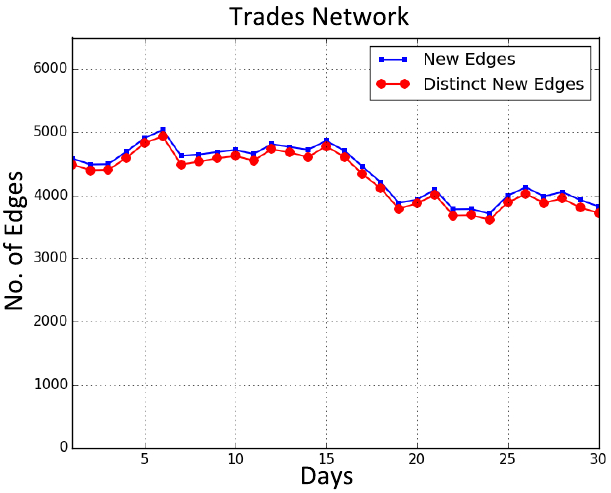}
\includegraphics[width=0.35\textwidth]{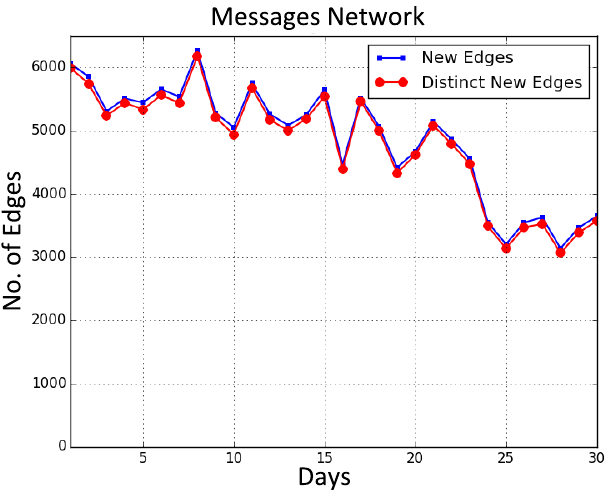}\\
  \caption{Dynamics of the Travian network (trades: left and messages: right). The blue line shows the new edges added, and the red line shows edges that did not exist in the previous snapshot.}
  \label{fig:travian-edges}
\end{figure}

For our analysis, we selected three datasets: player communication and economic networks from the Travian massively multiplayer online game~\cite{hajibagheri2015conflict}, the popular Enron email dataset~\cite{cohen2009enron}, and the co-authorship network from arXiv hep-th~\cite{soares2012time}.  Table~\ref{tab:network_stats} gives the network statistics for each of the datasets:

\begin{compactenum}
\item \textbf{Enron email dataset}~\cite{cohen2009enron}: This email network shows the evolution of the Enron company organizational structure over 24 months (January 2000 to December 2001).
\item \textbf{Travian MMOG}~\cite{hajibagheri2015conflict}:  We used the communication and trading networks of users playing the Travian massively multiplayer online game.  Travian is a browser-based, real-time strategy game in which the players compete to create the first civilization capable of constructing a Wonder of the World.  The experiments in this paper were conducted on a 30 day period in the middle of the Travian game cycle (a three month period). Figure~\ref{fig:travian-edges} indicates that Travian is a highly dynamic dataset, with over $90\%$ of the edges changing between snapshots.
\item \textbf{co-authorship network hep-th arXiv}~\cite{soares2012time}: This co-authorship network shows the evolution in co-authorship relationships extracted from the arXiv High Energy Physics (Theory) publication repository between 1991 and 2010.
\end{compactenum}

\begin{table*}
\renewcommand{\arraystretch}{1}
\caption{Dataset Summary}
\label{tab:network_stats}
\begin{center}
\begin{tabular}{lC{2.4cm}C{2.4cm}C{2.4cm}C{2cm}}
\toprule
\textbf{Data} &  
\textbf{Enron} &
\textbf{Travian (Messages)} & \textbf{Travian (Trades)} &
\textbf{hep-th}\\
\midrule
\textbf{No. of nodes} & 150 & 2,809 & 2,466  & 17,917\\
\textbf{Link (Class 1)} & 5,015 & 44,956 & 87,418  & 59,013\\
\textbf{No Link (Class 0)} & 17,485 & 7,845,525 & 5,993,738  & 320,959,876\\
\textbf{No. of snapshots} & 24 & 30 &30  & 20\\
\bottomrule
\end{tabular}
\end{center}
\end{table*}

\section{Method}

\name/ treats the link prediction problem as a supervised classification task, where each data point corresponds to a pair of vertices in the social network graph.  This is a typical binary classification task that could be addressed with a variety of classifiers; we use the Spark support vector machine (SVM) implementation.  All experiments were conducted using the default parameters of the Spark MLlib package: the SVM is defined with a polynomial kernel and a cost parameter of 1. Algorithms were implemented in Python and executed on a machine with Intel(R) Core i7 CPU and 24GB of RAM.  We have made our code and some example datasets available at: {\url{http://ial.eecs.ucf.edu/travian.php}}.

In order to produce a labeled dataset for supervised learning, we require timestamps for each node and edge to track the evolution of the social network over time.  We then consider the state of the network for two different time periods $t$ and $t'$ (with $t < t'$). The network information from time $t$ is used to predict new links which will be formed at time $t'$.  One of the most important challenges with the supervised link prediction approach is handling extreme class skewness. The number of possible links is quadratic in the number of vertices in a social network, however the number of actual edges is only a tiny fraction of this number, resulting in large class skewness. 

The most commonly used technique for coping with this problem is to balance the training dataset by using a small subset of the negative examples.  Rather than sampling the network, we both train and test with the original data distribution and reweight the misclassification penalties.  Let $G(V,A)$ be the social network of interest. Let $G[t]$ be the subgraph of $G$ containing the nodes and edges recorded at time $t$. In turn, let $G[t']$ be the subgraph of $G$ observed at time $t'$. In order to generate training examples, we considered all pairs of nodes in $G[t]$.  Even though this training paradigm is more computationally demanding it avoids the concern that the choice of sampling strategy is distorting the classifier performance~\cite{Lichtenwalter2010}.

Selecting the best feature set is often the most critical part of any machine learning implementation.  In this paper, we supplement the standard set of features extracted from the graph topology (described in the previous section), with features predicted by a set of time series.  Let $F_t(t = 1,...,T)$ be a time series with $T$ observations with $A_t$ defined as the observation at time $t$ and $F_{t+1}$ the time series forecast at time $t+1$.  First, we analyze the performance of the following time series forecasting models for generating features:
\begin{compactenum}
\item \textbf{Simple Mean}: The simple mean is the average of all available data:
	\[
F_{t+1} = \frac{A_{t}+A_{t-1}+...+A_{t-T}}{T}
\]
\item \textbf{Moving Average}: This method makes a prediction by taking the mean of the $n$ most recent observed values. The moving average forecast at time $t$ can be defined as:
	\[
F_{t+1} = \frac{A_{t}+A_{t-1}+...+A_{t-n}}{n}
\]
\item \textbf{Weighted Moving Average}: This method is similar to moving average but allows one period to be emphasized over others. The sum of weights must add to 100$\%$ or $1.00$:
	\[
F_{t+1} = \sum C_t A_t
\]
\item \textbf{Exponential Smoothing}: This model is one of the most frequently used time series methods because of its ease of use and minimal data requirements.  It only needs three pieces of data to start: last period's forecast ($F_t$), last period's actual value ($A_t$) and a value of smoothing coefficient,$\alpha$, between 0 and 1.0. If no last period forecast is available, we can simply average the last few periods:
	\[
F_{t+1} = \alpha A_t + (1-\alpha) F_t
\]
\end{compactenum}

\begin{figure}
  \centering 
  \subfloat[]{\includegraphics[width=0.3\textwidth]{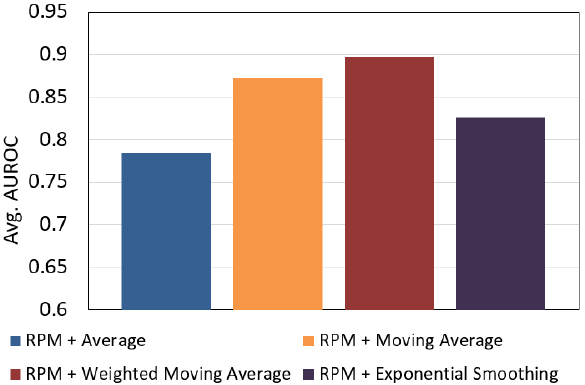}}
  \subfloat[]{\includegraphics[width=0.3\textwidth]{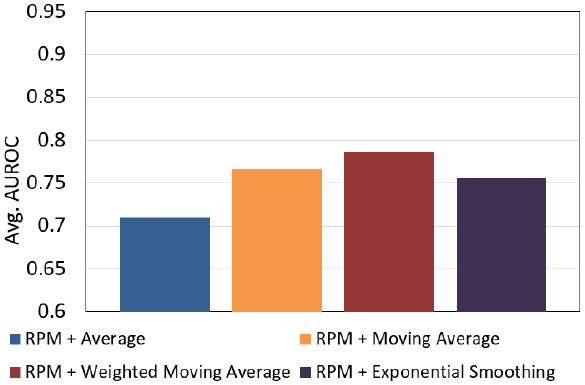}}
  \subfloat[]{\includegraphics[width=0.3\textwidth]{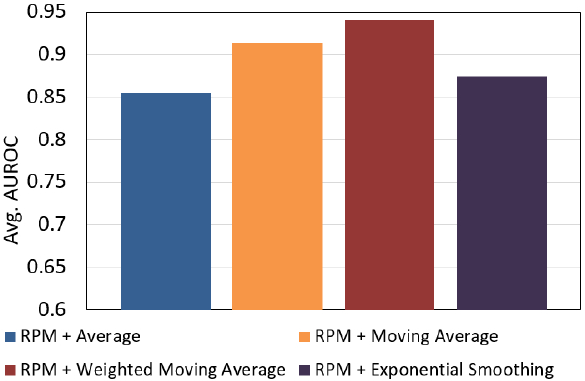}}
    \caption{Performance of \name/ using different forecasting models on (a) Travian Messages (b) hep-th (c) Enron.  Weighted Moving Average is consistently the best performer across all datasets and is used in \name/.}
    \label{fig:ts_performance}
\end{figure}

We identify which time series prediction model produces the best rate estimate, according to the AUROC performance of its \name/ variant.  Parameters of weighted moving average and exponential smoothing were tuned to maximize performance on the training dataset.  Figure~\ref{fig:ts_performance} shows that the best performing model was Weighted Moving Average with $n=3$ and parameters $C_1, C_2$ and $C_3$ set to 0.2,0.3, and 0.5 respectively. 

\subsection{Results}

Our evaluation measures receiver operating characteristic (ROC) curves for the different approaches. These curves show achievable true positive rates (TP) with respect to all false positive rates (FP) by varying the decision threshold on probability estimations or scores.  For all of our experiments, we report area under the ROC curve (AUROC), the scalar measure of the performance over all thresholds.  Since link prediction is highly imbalanced, straightforward accuracy measures are well known to be misleading; for example, in a sparse network, the trivial classifier that labels all samples as missing links can have a 99.99$\%$ accuracy. 

\begin{table*}[t]
  \caption{AUROC Performance}
\label{tab:auroc}
\begin{center}
  \bgroup
  \def\arraystretch{1.2}
  \begin{tabular*}{1\textwidth}{p{4cm}C{2.7cm}C{2.3cm}C{1.4cm}C{1.4cm}}
  	\hline
	Algorithms / Networks & Travian(Messages) & Travian(Trades) & Enron & hep-th\\ \hline
	 \textbf{\name/} & \textbf{0.8970} & \textbf{0.7859} & \textbf{0.9399} & \textbf{0.7834}  \\
	\textbf{Supervised-MA} &  0.8002 & 0.6143 & 0.8920 & 0.7542  \\
 \textbf{Supervised} &  0.7568 & 0.7603 & 0.8703 & 0.7051  \\
     \textbf{Common Neighbors} &  0.4968 & 0.5002 & 0.7419 & 0.5943 \\
     	 \textbf{Jaccard Coefficient} & 0.6482 & 0.4703 & 0.8369 & 0.5829 \\
	 \textbf{Preferential Attachment} & 0.5896 & 0.5441 & 0.8442 & 0.5165\\
	 \textbf{Adamic/Adar} & 0.5233 & 0.4962	 & 0.7430 & 0.6696\\
    \hline
  \end{tabular*}
  \egroup
 	\end{center}
\end{table*}

In all experiments, the algorithms were evaluated with stratified 10-fold cross-validation. For more reliable results, the cross-validation procedure was executed 10 times for each algorithm and dataset.  We benchmark our algorithm against \textbf{Supervised-MA}~\cite{soares2012time}.  Supervised-MA is a state of the art link prediction method that is similar to our method, in that it is supervised and uses moving average time series forecasting.  In contrast to \name/, Supervised-MA creates time series for the unsupervised metrics rather than the link formation rate itself.  \textbf{Supervised} is a baseline supervised classifier that uses the same unsupervised metrics as features without the time series prediction model.  As a point of reference, we also show the unsupervised performance of the individual topological metrics: 1) \textbf{Common Neighbors}, 2) \textbf{Preferential Attachment}, 3) \textbf{Jaccard Coefficient}, and 4) \textbf{Adamic-Adar}.  Table~\ref{tab:auroc} presents results for all methods on Travian (communication and trade), Enron, and hep-th networks.  Results for our proposed method are shown using bold numbers in the table; in all cases, \name/ outperforms the other approaches.  Two-tailed, paired t-tests across multiple network snapshots reveal that the RPM is significantly better ($p < 0.01$) on all four datasets when compared to Supervised-MA.

We discover that explicitly including the rate feature (estimated by a time series) is decisively better than the usage of time series to forecast topological metrics.   The rate forecast is useful for predicting the source node of future links, hence \name/ can focus its search on a smaller set of node pairs.   We believe a combination of topological metrics is useful for predicting the destination node, but that relying exclusively on the topological metrics, or their forecasts, is less discriminative.



\section{Related Work}

The performance of \name/ relies on three innovations: 1) explicit modeling of link formation rates at a node level, 2) the usage of multiple time series to leverage information from earlier snapshots, 3) training and testing with the full data distribution courtesy of the Spark fast cluster computing system.  Rate is an important concept in many generative network models, but its usage has been largely ignored within discriminative classification frameworks.  For instance, the stochastic actor-oriented model of network dynamics contains a network rate component that is governed by both the time period and the actors~\cite{snijders2010}.  \name/ does not attempt to create a general model of how the rate is affected by the properties of the actor (node), but instead predicts the link formation rate of each node with a time series.

Time series are useful because they enable us to track the predict future network dynamics, based on the past changes. Soares and Prud\^encio~\cite{soares2012time} investigated the use of time series within both supervised and unsupervised link prediction frameworks.  The core concept of their approach is that it is possible to predict the future values of topological metrics with time series; these values can either be used in an unsupervised fashion or combined in a supervised way with a classifier.  In this paper, we compare \name/ to the best performing version of their methods, Supervised-MA (Supervised learner with Moving Average predictor), that we reimplemented in Spark and evaluated using our full test/train distribution paradigm, rather than their original sampling method.  Predicting the rate directly was more discriminative that predicting the topological metrics.  We predict the rate of the source node's link formation using a time series, in contrast to Huang et al.~\cite{huang2009time} who used a univariate time series model to predict link probabilities between node pairs.  In our work, we use a supervised model to assign links, rather than relying on the time series alone.

Feature selection is especially critical to the performance of a supervised classifier. For co-authorship networks, Hasan et al.~\cite{hasan2006} identified three important categories of classification features: 1) proximity (for comparing nodes) 2) aggregated (for summing features across nodes), and 3) topological (network-based).  In our work, we only use network-based features, since those are the easiest to generalize across different types of networks; both proximity and aggregated features require more feature engineering to transfer to different datasets.  Wang and Sukthankar~\cite{LNCS2014-Xi} promoted the importance of social features in both supervised and unsupervised link prediction; social features aim to express the community membership of nodes and can be used to construct alternate distance metrics.  However we believe that rate generalizes better across different types of dynamic networks; moreover it can be easily combined with dataset-specific feature sets.

\section{Conclusion and Future Work}
In this paper, we introduce a new supervised link prediction method, \name/ (Rate Prediction Model), that uses time series to predict the rate of link formation.  By accurately identifying the most active individuals in the social network, \name/ achieves statistically significant improvements over related link prediction methods.  Unlike the preferential attachment metric which identifies active individuals based on the degree measure of a single snapshot, \name/ measures time-sliced network structure and finds individuals whose influence is rapidly rising.  Our experiments were performed on networks created by a variety of social processes, such as communication, collaboration, and trading; they show that the rate of link generation varies with the type of network.  In future work, we plan to extend this method to do simultaneously link prediction on different layers of multiplex networks, such as Travian, by modeling the relative rate difference between network layers.

\section{Acknowledgments}
Research at University of Central Florida was supported with an internal Reach for the Stars award.  Sandia National Laboratories is a  multi-program laboratory managed and operated by Sandia Corporation, a wholly owned subsidiary of Lockheed Martin Corporation, for the U.S. Department of Energy's National Nuclear Security Administration under contract DE-AC04-94AL85000. 

\bibliographystyle{splncs}
\bibliography{references}
\end{document}